\documentclass[a4paper,fleqn,usenatbib]{mnras}

\usepackage{newtxtext,newtxmath}

\usepackage[T1]{fontenc}
\usepackage{ae,aecompl}

\usepackage{graphicx}	
\usepackage{amsmath}	
\usepackage{amssymb}	

\usepackage{hyperref}	
\hypersetup{colorlinks=true,linkcolor=blue,citecolor=blue,filecolor=blue,urlcolor=blue}
\def\TZO{T\.ZO}
\def\TZOs{T\.ZOs}
\def\ga{\mathrel{\hbox{\rlap{\hbox{\lower4pt\hbox{$\sim$}}}\hbox{$>$}}}}
\def\la{\mathrel{\hbox{\rlap{\hbox{\lower4pt\hbox{$\sim$}}}\hbox{$<$}}}}

\def\lbol{$L$\mbox{$_{\rm bol}$}}

\def\teff{$T_{\rm eff}$}

\def\logl{$\log (L/L_\odot)$}

\def\teff{$T_{\rm eff}$}
\def\one{{\sc{i}}}
\def\logl{$\log(L/{\rm L}_\odot)$}

\defcitealias{levesque2014discovery}{L14}

\title[Re-evaluation of \TZO\ candidate HV 2112]{A critical re-evaluation of the Thorne-\.Zytkow object candidate HV~2112}

\author[E. R. Beasor et al.]{
Emma R. Beasor$^{1}$\thanks{E-mail:e.beasor@2010.ljmu.ac.uk},
 Ben Davies$^{1}$,
Ivan Cabrera-Ziri$^{2}\thanks{Hubble Fellow}$,
\& Georgia Hurst$^{1}$
\\
$^{1}$Astrophysics Research Institute, Liverpool John Moores University, Liverpool,  L3 5RF, UK\\
$^{2}$Harvard-Smithsonian Center for Astrophysics, 60 Garden Street, Cambridge, MA 02138, USA}
\date{Accepted XXX. Received YYY; in original form ZZZ}

\pubyear{2018}

\begin{document}
\label{firstpage}
\pagerange{\pageref{firstpage}--\pageref{lastpage}}
\maketitle

\begin{abstract}

It has been argued in the literature that the star HV~2112 in the Small Magellanic Cloud is the first known example of a \TZO, a Red Supergiant with a degenerate neutron core. This claim is based on the star having a high luminosity (\logl $\ga$ 5), an extremely cool effective temperature, and a surface enriched in  in lithium, calcium and various $irp$-process elements. In this paper we re-examine this evidence, and present new measurements of the stellar properties. By compiling archival photometry from blue to mid-IR for HV~2112 and integrating under its spectral energy distribution we find a bolometric luminosity in the range of \logl=4.70-4.91, lower than that found in previous work and comparable to bright asymptotic giant branch (AGB) stars.  We compare a VLT+XSHOOTER spectrum of HV~2112 to other late type, luminous SMC stars, finding no evidence for enhancements in Rb, Ca  or K, though there does seem to be an enrichment in Li. We therefore conclude that a much more likely explanation for HV~2112 is that it is an intermediate mass($\sim$5M$_\odot$) AGB star. However, from our sample of comparison stars we identify a new \TZO\ candidate, HV~11417, which seems to be enriched in Rb but for which we cannot determine a Li abundance.
\end{abstract}

\begin{keywords}
stars: abundances -- stars: individual: HV 2112 -- stars: peculiar -- stars: supergiants
\end{keywords}



\section{Introduction}
Thorne-\.Zytkow objects \citep[\TZOs,][]{thorne1975red,thorne1977stars} are a theoretical class of stellar object in which a neutron star sits within the diffuse envelope of a supergiant star. There are various suggested formation channels, most of which involve the merger of a binary system \citep[e.g.][]{thorne1977stars,taam1978double}. These objects are thought to be extremely rare, with as few as 20-200 \TZOs\ predicted to exist in the Galaxy at present \citep{podsiadlowski1995evolution}, though some authors have doubted whether such an object could survive the merger with the envelope intact \citep{papish2015ejecting}.

Identifying a \TZO\ is challenging as it is expected that they would be virtually indistinguishable from normal red supergiants (RSGs) close to the Hayashi limit \citep{thorne1977stars}. They are expected to be highly luminous (\logl$\ga$5), have cool effective temperatures (\teff, $\la$3000K) and enriched in interrupted rapid protion ($irp$) processed elements, such as rubidium and molybdenum. They are also expected to be enriched in lithium as helium burning takes place at the base of the convective zone\citep{podsiadlowski1995evolution}. These elements are not expected to be enhanced in normal RSGs, so identifying a \TZO\ candidate relies strongly on the ability to measure the abundances of these heavy elements\footnote{In the original papers \citep{thorne1975red,thorne1977stars} a low mass \TZO\ was also proposed. These objects would appear cooler than typical RSGs and potentially be dynamically unstable, but would not show the abundance anomalies or higher luminosities expected for high mass \TZOs\ making them virtually indistinguishable from AGB stars.}.

HV~2112 was speculated to be a \TZO\ in \cite{levesque2014discovery} (hereafter L14) and has now been confirmed as member of the Small Magellanic Cloud (SMC) from its proper motion in Gaia-DR2 \citep{gaiadr2,mcmillan2018gaia} after a brief controversy over its distance \citep{maccarone2016large}. The claim by L14 was based on three pieces of evidence. Firstly, L14 determined a high luminosity (\logl = 5.02) for HV~2112, above that expected for asymptotic giant branch (AGB) stars. Secondly, it was argued that the star's surface was enriched in molybdenum, rubidium, lithium and, unexpectedly, calcium. Finally, HV 2112 has a late spectral type in comparison to the SMC average for RSGs. 

 \cite{tout2014hv2112} discussed in depth the possibility of whether or not HV~2112 could instead be a super asymptotic giant branch star (SAGB), a class of star that is also expected to exhibit enhanced lithium and have a high \lbol. A finely tuned model was required to explain the $irp$ processed elements. Interestingly, \citeauthor{tout2014hv2112} showed that an overabundance in Ca is difficult to explain with either the \TZO\ or AGB scenario.

In this paper we re-evaluate the above evidence in support of the \TZO\ classification for HV~2112. In the following section we discuss the surface abundances of HV~2112, by comparing the star's spectrum to other late type SMC stars, thee effective temperature of the star, and we re-appraise the luminosity of HV~2112 using archival data. 

\section{A re-evaluation of the evidence}
\subsection{Chemical abundances}
It is predicted that \TZOs\ would have unusual abundance patterns due to the extreme temperatures at the surface of the NS, showing enhancement in a number of heavy elements due to the $irp$-process  \citep[e.g. Li\,\one, Rb\,\one, Mo\,\one, ][]{biehle1991high,cannon1993massive,podsiadlowski1995evolution}. It was claimed in L14 that HV2112 displayed overabundances in Rb\,\one, Mo\,\one, Li\,\one\ and Ca\,\one. These conclusions were not based on quantitative abundance analysis using model atmospheres. Instead, they determined pseudo equivalent widths\footnote{It is not possible to measure a strict equivalent width due to the high opacity in the optical spectra of cool stars. This is discussed further in Section 2.1.1. } of various diagnostic lines, and compared certain line ratios to those of a sample of RSGs in the SMC. Any line ratio for HV~2112 that lay outside the 3$\sigma$ limit of the comparison sample was considered to be indicative of an anomalous abundance ratio. 

From the results presented in their Fig. 1, \citetalias{levesque2014discovery} argue for an overabundance of rubidium (from the Rb\,\one\ $\lambda\lambda$7800.23 / Ni\,\one\  $\lambda\lambda$6707.97 ratio), lithium (from the Li\,\one\ $\lambda\lambda$6707.97/Ca\,\one\ $\lambda\lambda$6572.78 and Li\,\one\ $\lambda\lambda$6707.97/K\,\one\ $\lambda\lambda$7698.97 ratios), calcium (from the Ca\,\one\ $\lambda\lambda$6572.78/Fe\,\one\ $\lambda\lambda$7802.47 ratio) and molybdenum (from the Mo\,\one\ $\lambda\lambda$5570.40/Fe\,\one\ $\lambda\lambda$5569.62 ratio). However, attributing these line ratios to the abundance of the elements poses some problems. Firstly, the Rb/Ni ratio suggests an $over$ abundance of Rb, while the Rb/Fe ratio is normal. However, from the same figure the Ni/Fe ratio also appears to be within the normal 3$\sigma$ range. If we attribute these line ratios directly to abundance ratios, it is not possible to find a self-consistent explanation. There is a similar problem for Li, Ca and K. The Li/Ca and Li/K ratios could be interpreted as an over abundance in Li, while the Ca/Fe ratio appears to suggest an over abundance in Ca. However, the K/Ca ratio implies no over abundance in Ca, and appears normal. 

Though we cannot explain the line strength ratios, it could be due to L14's comparison stars having very different temperatures and log(g) values, and it is likely that the lines in question are sensitive to such factors. There is also the issue that these lines are heavily blanketed by TiO, which  is highly sensitive to $T_{\rm eff}$ and atmospheric structure (see Section 2.2). For these reasons we argue that it is not possible to draw conclusions about the abundances of the heavy elements from the L14 analysis alone. 

\subsubsection{Comparative Spectroscopy}
In this section, we attempt to circumvent the issues described in the previous section by comparing a spectrum of HV2112 to those of other stars in the SMC with similar luminosities and spectral types. To this end we have obtained an VLT+XSHOOTER spectrum of HV~2112 from the VLT archive (PI: C. Worley, ID: 096.D-0911(A)) as well as data from the X-Shooter Library \citep{Chen14}. The sample of stars is listed in Table \ref{tab:sample}, along with their basic properties. The comparison stars were chosen for their late spectral types and SMC membership. Most are thought to be AGB stars from their moderately high luminosities ($\log(L/{\rm L}_\odot) \ga 4$), late spectral types and large amplitude photometric and spectroscopic variability (often $\ga$1 mag in $V$ over timescales of 100s of days.). 

\begin{figure}
\begin{center}
\includegraphics[width=8.5cm]{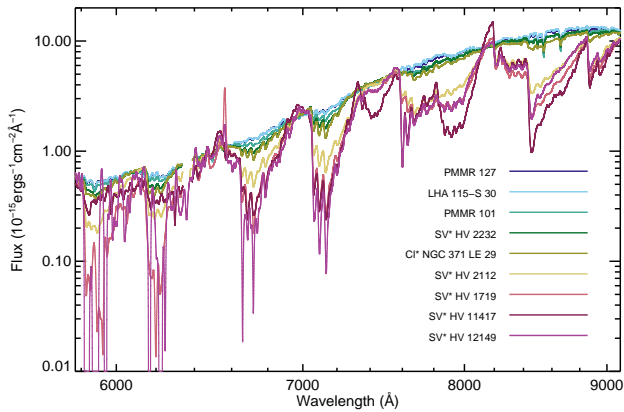}
\caption{Spectral energy distributions of the stars in our sample. The spectrum of HV~2112 is flux-calibrated, whereas the other spectra are normalized according to the flux of HV2112 in the relatively line-free regions in between the TiO bands. }
\label{fig:seds}
\end{center}
\end{figure}

The optical spectral energy distributions (SEDs) of the stars are shown in Fig.\ \ref{fig:seds}. The spectra have been normalized to the apparently line-free spectral regions of HV~2112 to better illustrate the progression in absorption strengths with spectral type. To make a quantitative measurement analogous to spectral type we define the `TiO index' to be the ratio of the fluxes either side of the TiO bandhead at 7020\AA. 

In addition, for all stars in the sample we have measured the strengths of all lines studied by L14, specifically Rb\,\one\ $\lambda$7800.23, Li\,\one\ $\lambda$6707.97, Ca\,\one\ $\lambda$6572.78 and K\,\one\ $\lambda$7698.97. The high opacity in the optical spectra of cool stars prevents the measurement of a strict equivalent width since it is not possible to know where the true continuum is located. Instead we measured the pseudo-equivalent widths (pEWs), where the pseudo-continuum is defined to be the linear interpolation of flux-maxima either side of the line-centre, similar to L14. We use Gaussian profile fitting to the absorption profile to determine the pEW of the spectral line, and we estimate uncertainties on pEW by moving the placement of pseudo-continuum blue/red by one pixel and remeasuring the line strength.  

\begin{figure}
\begin{center}
\includegraphics[width=8.5cm]{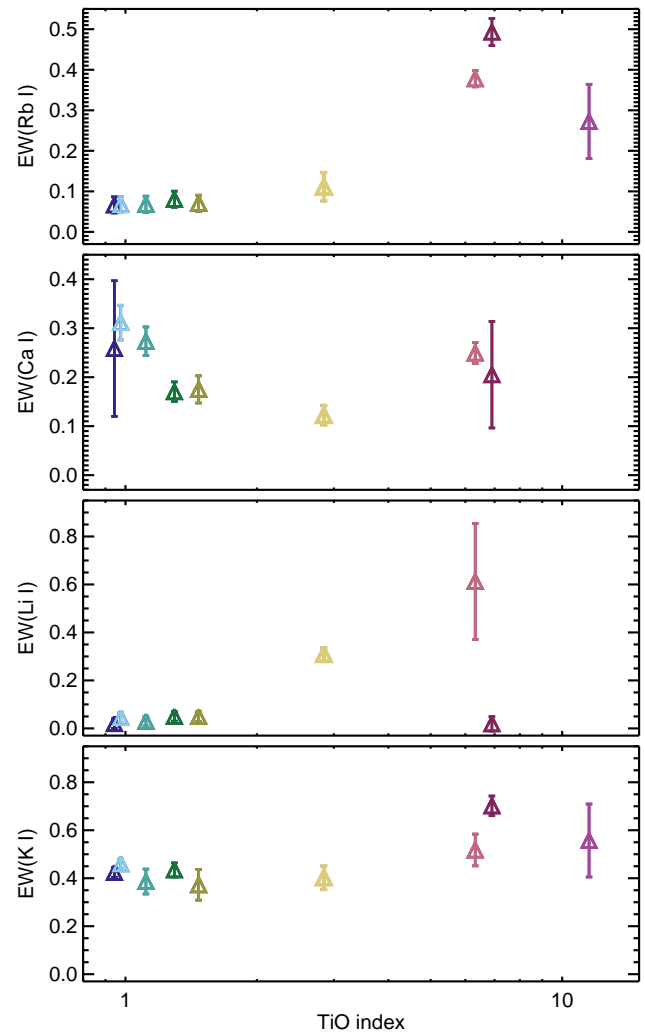}
\caption{Pseudo equivalent widths (pEWs) of the four spectral lines studied in L14, as a function of TiO absorption index. HV~2112 is denoted by the yellow triangle. For the star HV~12149 (the star with the highest TiO index), reliable measurements of the Ca\,\one\ and Li\,\one\ lines could not be made due to the low signal-to-noise of the spectrum at those wavelengths.}
\label{fig:elem}
\end{center}
\end{figure}

In Fig.\ \ref{fig:elem} we plot the pEWs of each line studied as a function of the TiO index. HV~2112 is plotted as the large filled triangle. Though HV~2112 is somewhat of an outlier in terms of its spectral type when compared to typical RSGs in the SMC, the absorption in the lines of of Rb\,\one, Ca\,\one\ and K\,\one\ do not stand out as remarkable in comparison to the other late-type SMC stars. Indeed, they seem to be consistent with the very shallow trend with spectral type. The exception is the Li\,\one\ line, where we see an apparent dichotomy of either effectively no Li absorption or a pEW$>$0.3\AA. Under the simple assumption that line strengths at a given spectral type reflect the abundance of the element, this suggests that HV~2112 may be Li-enriched. By contrast, we see no evidence for enhancement in any of the elements Rb, Ca or K. 

One star in our sample which does potentially show Rb enhancement is HV~11417, which can be seen in Fig.\ \ref{fig:elem} as the star with the highest Rb\,\one\ pEW. The star shows no evidence of Li enhancement, though the TiO bands are so strong in this star's spectrum that they may completely overwhelm the Li\,\one\ $\lambda$6707.97 line. Combined with its high luminosity (\logl=4.92), we consider this star to be a TZO candidate until its Li abundance can be measured. Such a measurement may be possible when the star is in a phase of variability where it has an earlier spectral type and the  Li\,\one\ line is more clearly detectable.

\begin{table*}
\caption{Comparison stars used in this work. Columns show the object name, co-ordinates, $K$-band magnitude from 2MASS, bolometric luminosity$^\dagger$, variability amplitude and period/timescale for variability, waveband in which the variability was measured and the associated reference.  }
\begin{center}
\begin{tabular}{lccccccc}
\hline \hline
Name & RA DEC (J2000) & $K_S$ & \logl & $\Delta m_\lambda$ & $P$ (days) & $\lambda$ & ref \\
\hline 
        SV* HV 1719 &  00 57 14.5  -73 01 21.3 & 9.80 & 4.34 & 2.49 & 542 & I  & \cite{soszynski2009optical}  \\
       SV* HV 12149 &  00 58 50.2  -72 18 35.6 & 8.61 & 4.81 & 2.35 & 769 & I  & \cite{soszynski2009optical}  \\
           PMMR 101 &  00 59 35.0  -72 04 06.6 & 8.35 & 4.92 & 0.12 & 394 & V & \cite{groenewegen2018luminosities}  \\
       LHA 115-S 30 &  01 00 41.5  -72 10 37.0 & 7.96 & 5.07 & 0.61 & 351 & V & \cite{watson2006international}  \\
       SV* HV~11417 &  01 00 48.2  -72 51 02.1 & 8.45 & 4.87 & 1.86 & 1092 & I  & \cite{soszynski2009optical}  \\
           PMMR 127 &  01 01 54.2  -71 52 18.7 & 8.69 & 4.78 &      &      &    &    \\
  Cl* NGC 371 LE 29 &  01 03 02.5  -72 01 53.1 & 8.62 & 4.81 & 0.84 & 543 & I  & \cite{soszynski2009optical}  \\
        SV* HV~2112 &  01 10 03.9  -72 36 52.6 & 8.72 & 4.70 - 4.91 & 1.27 & 603 & I  & \cite{soszynski2009optical}  \\
        SV* HV 2232 &  01 30 34.0  -73 18 41.7 & 8.60 & 4.82 & 0.47 & 506 & V & \cite{groenewegen2018luminosities}  \\
        
\hline
\multicolumn{8}{l}{$^\dagger$The luminosity here is calculated using $K_{\rm S}$-band photometry and BC$_{K}$=3 except for HV 2112 where the luminosity was determined} \\
\multicolumn{8}{l}{using the method described in Section 2.3.} \\

\end{tabular}
\end{center}
\label{tab:sample}
\end{table*}%

\subsection{Effective temperature}
One key attribute of \TZOs\  is that they are expected to be cooler than the average RSG \citep[][]{thorne1977stars}.  To estimate the temperature of HV~2112, L14 fit MARCS stellar atmosphere models \citep{gustafsson2008grid} to the optical SED of HV~2112, deriving a $T_{\rm eff}$ of 3450K. In Figure 3 of L14, it can be seen that there is a discrepancy between the spectrum of HV~2112 and the best fit MARCS model at near-UV and a near-IR wavelengths (an excess and deficiency, respectively). The authors attribute this to circumstellar dust and strong mass loss, also described in \cite{levesque2005physical,levesque2006effective}. It has also been argued that 3-D model atmospheres produce greater TiO absorption at a given effective temperature in comparison to 1-D models \citep{davies2013temperatures}. Currently no models capable of reproducing the whole spectrum of RSGs, though 3-D hydrodynamical models show promise \citep{chiavassa2011radiative}. For this reason, we conservatively estimate the temperature of HV~2112 using Wien's displacement law and the position of the flux peak, somewhere between I and J band. From this we estimate a temperature in the range of 2500-3750K.

\subsection{Luminosity}
In L14, the luminosity of HV~2112 was estimated to be \logl = 5.02. This was calculated by correcting the V-band magnitude using the bolometric correction (BC) from \cite{levesque2006effective}, a distance modulus of 18.9 and an extinction of $A_{\rm V}$$\sim$0.4. However, as the BC is dependent on \teff\ estimates from \cite{levesque2006effective}, which as previously discussed (see Section 2.2) may be overestimated, the BC itself may not be reliable.

A less model dependent method to estimate \lbol\  is to adopt the method described in \cite{davies2018humphreys}, in which the SED is integrated under from the blue to the mid-IR. This method does not rely on uncertain BCs to determine \lbol. Our only assumptions are that the emitted flux is spherically symmetric and any flux lost to circumstellar extinction in the optical is re-radiated at longer wavelengths. 

We compiled photometry from OGLE \citep{udalski1992optical}, ASAS \citep{pojmanski1997all}, DENIS \citep{cioni2000denis}, 2MASS \citep{skrutskie2006two}, Spitzer \citep{werner2004spitzer} and WISE \citep{wright2010wide}. This photometry was de-reddened according to the SMC extinction law of \cite{gordon2003quantative} for an $A_{\rm V}$ value of 0.56$\pm$0.26, found by interpolating the extinction maps of \cite{zaritsky2002dust} to the position of HV2112 \citep[see ][]{davies2018luminosities}. To account for any missing flux at short wavelengths, a black body spectrum of $T_{\rm eff}$ = 3000K was matched to the B-band flux\footnote{The total luminosity contribution of this was small, approximately 0.02 dex. We also varied the effective temperature of the black body by $\pm$1000K and it changed the value of luminosity by $<<$ 0.01 dex.}. We then integrated under the spectrum using {\tt IDL} routine {\tt int}$\_${\tt tabulated} and corrected for the distance to the SMC.

One factor that might impact our determined \lbol\ is HV~2112's variability. The star is known to be a long period variable, and has been photometrically monitored by OGLE and ASAS. Specifically, HV~2112 is variable in V-band by an amplitude of 2.2 mag (OGLE), I-band by an amplitude of 2.1 mag (ASAS). In addition, HV~2112 has been observed twice by the DENIS survey, implying variability with an amplitude of at least 0.3 mag at J and of 0.13 mag at K. As these observations are not contemporaneous, we do not know how the colour evolves with the observed variability.  It is therefore unclear if the variability is due to the star changing in \lbol, \teff\ or both. To find an upper and lower limit to HV~2112's luminosity during variability we will assume that BC does not change, see Fig. 3. By assuming the observed variability is due solely to changes in luminosity, we find a range of \logl\ = 4.70 - 4.91. The variability in luminosity is low, 0.21 dex, as despite the changing brightness at V and I bands most of the star's flux is emitted at $\lambda$>1$\micron$, because of this the luminosity does not change by a large amount. 

Figure \ref{fig:HRD} shows the luminosity and temperature range for HV~2112 on a Hertzsprung-Russel diagram (HRD). We have overplotted stellar evolution models from  BPASS \citep{eldridge2009spectral} to demonstrate how the luminosity of a star changes with evolution. From these models, we cannot rule out an intermediate mass star ($\sim$5M$_\odot$) in the AGB phase, as thermal pulses during this time are predicted to increase the luminosity to \logl $\sim$5. This behaviour is also shown by MIST \citep{dotter2016mesa} and Padova \citep{salasnich1999evolution} evolutionary tracks.

\begin{figure}
\begin{center}
  \caption{ SED for HV~2112. The photometry is shown by black circles. The red line shows the flux at the peak of variability, while the purple line shows the flux at the minimum of variability, assuming the bolometric correction of the star does not change. The shaded green section shows the possible range of fluxes. }

  \label{fig:SED}
     \includegraphics[width=8.5cm]{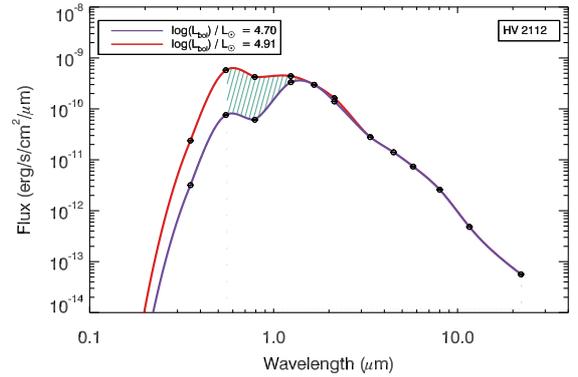}
\end{center}
\end{figure}

\begin{figure}
\begin{center}
  \caption{Position of HV~2112 on a HR diagram with BPASS models over plotted. The location of HV~2112 is consistent with that of a 5M$_\odot$ star in the AGB phase. }
  
    \label{fig:HRD}
     \includegraphics[width=8.5cm]{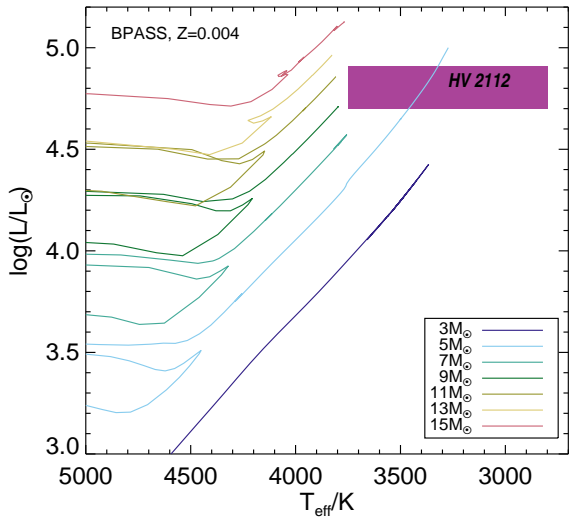}
\end{center}
\end{figure}

\section{Discussion and summary}
We have re-examined the evidence for the \TZO\ candidate HV2112. Below we summarise our results:

\begin{itemize}
\item We have compared archival spectra of HV~2112 to that of other SMC stars with similarly late spectral types. We find that the line strengths of K\,\one, Ca\,\one\ and Rb\,\one\ are normal with respect to the comparison stars, suggesting no evidence of enhancement of any of these elements. The only exception to this is Li which may be enriched.

\item Using archival photometry from 0.4 -25$\micron$ and integrating under the SED we find a bolometric luminosity for HV~2112 of \logl=4.72, with an upper limit of \logl=4.91. This limit comes from the star's variability under the conservative assumption that the bolometric correction does not change. This luminosity is lower than previously suggested, and is consistent with predictions for 5M$_\odot$ AGB stars which can have luminosities of \logl $\ga$ 5 \cite[e.g. ][]{eldridge2009spectral}.  

\item While the precise temperature of HV~2112 remains unknown, it is clear that it is a cool star ($T_{\rm eff}$$\sim$3000K).

\end{itemize}
Given the downward revision in luminosity and re-evaluation of the surface abundances in HV~2112, we argue that HV~2112 does not meet the criteria to be considered a \TZO candidate. Instead, it is likely a thermally pulsing AGB star. 

However, we also identify another \TZO\ candidate in HV~11417. This object has a high luminosity (\logl = 4.92) and seemingly strong Rb\,\one\ line. The star has a very late spectral type, so much so that the deep TiO absorption bands do not allow us to measure the strengths of the $\lambda$6707.97 Li line. We suggest that HV 11417 is a more promising candidate \TZO, pending a measurement of the Li abundance. Such a measurement could be made  when the star is in an earlier spectral-type phase of its variability. Spectral monitoring may allow observations at an earlier spectral type, when the Li line could be seen.

\section*{Acknowledgements}
The authors would like to thank the referee for useful comments which helped improve the paper and Maurizio Salaris for useful discussion. Support for this work was provided by NASA through Hubble Fellowship grant HST-HF2-51387.001-A awarded by the Space Telescope Science Institute, which is operated by the Association of Universities for Research in Astronomy, Inc., for NASA, under contract NAS5-26555.



\bibliographystyle{mnras}
\bibliography{references} 




\bsp	
\label{lastpage}
\end{document}